\documentstyle[twoside,fleqn,espcrc2]{article}
\input epsf

\title{Can we do better than Hybrid Monte Carlo in Lattice QCD  ?}

\author{M.E.~Berbenni-Bitsch$^{\rm a}$,
        A.P.~Gottlob$^{\rm a}$, 
        S.~Meyer\address{Fachbereich Physik - Theoretische Physik,
        Universit\"at Kaiserslautern,\\
        D-67663 Kaiserslautern, Germany}\thanks{ Speaker at the 
        International Symposium on Lattice
        Field Theory, 4-8 June 1996, St. Louis,Mo, USA. This
        work was supported by DFG under grant Me 567/5-3 }
        and M.~P\"utz$^{\rm a}$\thanks{Present address: MPI 
        f\"ur Polymerforschung, Ackermannweg 10, 
        D-55021 Mainz, Germany}}

\begin{document}

\begin{abstract}

  The Hybrid Monte Carlo algorithm for the simulation of QCD with
  dynamical staggered fermions is compared with Kramers equation
  algorithm. We find substantially different autocorrelation times
  for local and nonlocal observables. The calculations
  have been performed on the parallel computer CRAY T3D.

\end{abstract}

\maketitle

\section{INTRODUCTION}

Overcoming critical slowing down in Monte Carlo simulations of lattice
field theories is imperative to approach the continuum limit. Collective
mode algorithms 
have improved the
quality of numerical studies for systems with bosonic degrees of freedom
substantially during the last years. Besides many other things the
important role played by autocorrelation functions 
has been recognized and in particular strong
finite size effects present in the decay of  autocorrelation
functions have been observed. 

On the other hand it is fair to say, that the improvements of fermionic
simulation algorithms for lattice field theories since the introduction
of the Hybrid Monte Carlo (HMC)  algorithm
\cite{HMC1} ten years ago have not
been overwhelming. The theoretical analysis of the HMC algorithm has been
possible in the free field case \cite{HMC2}
while the practical analysis to compare
the performance of variants of the HMC algorithm and other dynamical
fermion algorithms for lattice QCD has only be done for Wilson fermions
\cite{JA95,JA96}.

The work reported here is a first step to obtain reliable estimates
for integrated autocorrelation times of different operators in lattice
QCD with dynamical staggered fermions using two different
simulation algorithms.

\section{ AUTOCORRELATION FUNCTIONS}

In lattice quantum field theory one evaluates expectation values 
of observables $A(\phi)$
\begin{displaymath}
  \langle A(\phi) \rangle =
    \frac{1}{Z} \int\prod\limits_{x\in\Lambda} d\phi_x \, e^{-S(\phi)} A(\phi)
\end{displaymath}
with $S(\phi)$ the action and $\phi_x$ a field on a lattice $\Lambda$
by a dynamic Monte Carlo algorithm.

The {\it dynamics} of the numerical algorithm is a
stochastic process which is realized on a computer by a Markov process. The
transition probability matrix $P(\phi\rightarrow\phi')$ leaves the
equilibrium distribution invariant.

{\it Detailed balance}
is equivalent to the selfadjointness of $P$ as an operator on
$L^2(\mu)$ with real spectrum $(\lambda_{min}, \lambda_{max})$, and the
stationarity condition 
\begin{displaymath}
  \int\prod\limits_{x\in\Lambda} d\phi_x \, e^{-S(\phi)}
                        \, P(\phi\rightarrow\phi') =
  {\bf 1} \cdot e^{-S(\phi')}.
\end{displaymath}
may be read as an eigenvalue equation with
eigenvalue $\lambda = 1$ and eigenvector $e^{-S(\phi')}$.
All other eigenvalues $\lambda$ have the upper bound 1.

The exponential autocorrelation time $\tau_{exp}$ parameterizes the gap
between $\lambda=1$ and the subleading (unwanted) modes.
The {\it expected error} $\sigma_A$ of an observable $A(\phi)$ is
\begin{eqnarray*}
  \sigma_A^2 & = & \frac{1}{n}\sum\limits_{t=-(n-1)}^{n-1}
                   \left(1-\frac{|t|}{n}\right)
                   C_{AA}(t) \\
       & \approx & \frac{1}{n} 2 \tau_{int,A} \cdot C_{AA}(0), \qquad
                   \mbox{for } n \gg \tau_{exp}
\end{eqnarray*}
with the autocorrelation function
\begin{eqnarray*}
  C_{AA}(t) & = & \langle (A{(i)} - \bar A) (A{(i+t)} - \bar A) \rangle 
\end{eqnarray*}
and the integrated autocorrelation time
\begin{displaymath}
  \tau_{int,A} =
       \frac{1}{2}\sum\limits_{t=-\infty}^\infty \frac{C_{AA}(t)}{C_{AA}(0)}.
\end{displaymath}
 
The performance of dynamic Monte Carlo algorithms for a finite system
with linear size $L$ and correlation length $\xi$ is described by
{\it empirical dynamical scaling laws}
\begin{displaymath}
  \tau_{int,A} \sim \min (L,\xi)^{z_{int,A}}.
\end{displaymath}

\section{GENERALIZED HMC ALGORITHM }

Introduce a set of ``fictitious momenta'' $\pi$ and a Hamiltonian
\begin{displaymath}
  H(\phi,\pi) = \frac{1}{2}\pi^2 + S(\phi)
\end{displaymath}
then the HMC       alternates two Markov steps: {\it momentum refreshment}
and {\it Monte Carlo molecular dynamics} which contains molecular dynamics
and a global Metropolis step to correct for the discretization errors in
Hamilton's equations.

The first step is momentum refreshment, where
momenta $\pi$ are replaced by new values chosen at random from a Gaussian
distribution.

Horowitz \cite{HO91} suggested a generalized momentum
 refreshment (Kramers equation)
by adding a Gaussian noise
\begin{displaymath}
\pi^\prime = e^{-\gamma \delta\tau}\cdot \pi  +  
 \sqrt{1 - e^{-2\gamma \delta\tau}} \cdot \eta
\end{displaymath}
with
\begin{displaymath}
   P(\eta) = \frac{1}{Z}e^{-\eta^2/2}
\end{displaymath}
where $0 \leq \gamma \leq \infty $. For $\gamma = \infty$ one obtains
full momentum refreshment, while $\gamma = 0$ gives exact 
generalization of the second order Langevin algorithm.
Detailed balance will be satisfied if the new phase space configuration
is accepted with probability
\begin{displaymath}
  P\left[(\phi,\pi)\rightarrow (\phi',\pi') \right]= 
   \min (1, e^{H(\phi,\pi) - H(\phi',\pi')}).
\end{displaymath}

\section {NUMERICAL RESULTS}

All our results are for lattice QCD with four flavours of dynamical
staggered fermions and gauge group $SU(2)$.

In the HMC algorithm there are two free parameters:
the integration step size $\delta\tau$ and the
trajectory length $\tau_0 = n\;\delta\tau$, where $n$ is the number
of integration steps.
$\delta\tau$ must be adjusted to keep $P_{acc}$ reasonably large.
Optimal choice for $\tau_0$ is less clear. In the {\it Gaussian model} with
$L \gg \xi$ the dynamical critical exponent $z$ is $z=2$ for 
$\tau_0 = \mbox{constant}$, and $z = 1$ if $\tau_0 \propto\xi$.
For constant $\tau_0$ the dynamical critical exponent $z$ becomes $z=2$
for $\xi \gg L$.

For dynamical fermions with $m=0.1$ we measured autocorrelation times
for the plaquette, Polyakov loop, and the chiral order parameter
with trajectory length $1/4 \le \tau_0 \le 2 $ on various lattices
with sample sizes of~$\approx 1000\,\tau_{ int}$.

\begin{figure}[htb]
\vspace{9pt}
\epsfxsize 75mm
\epsfbox{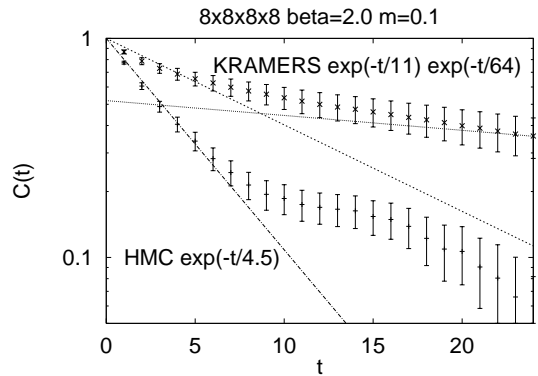}
\caption{Autocorrelation functions of the pla\-quette data.}
\label{fig:largenenough}
\end{figure}

In Fig. 1 we show the autocorrelation function of the plaquette 
on a lattice of size $8^4$ for both algorithms.
For the Kramers equation algorithm $k=12$ (see below) and 
on the horizontal axis $t$ is
given in units of molecular dynamics time. The straight lines are fixed 
exponentials to guide the eye.

For the Kramers equation algorithms the second exponential in the decay
of the autocorrelation function seemed to be stronger coupled. 
This observation has been found for  local as well as nonlocal
observables. 

\subsection{Generalized HMC algorithm}

Through the generalized momentum refreshment a new 
tunable parameter $\gamma$ is introduced.

Detailed balance requires that the momenta must have their sign flipped
after every rejected step.

The Monte Carlo molecular dynamics step can be performed $k$ times.
For dynamical fermions with $m=0.1$ we measure autocorrelation times
$\tau_{int,A}$ for the plaquette, Polyakov loop and the chiral order parameter
with $0.03 \le \delta\tau \le 0.12$, and  $k = 4,8$ and $12$.

\begin{figure}[htb]
\epsfxsize 75mm
\epsfbox{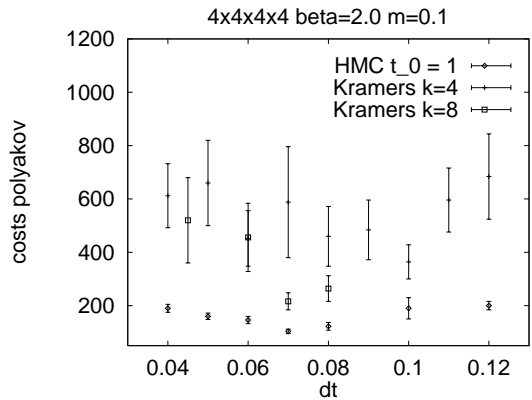}
\caption{Comparison of the HMC and Kramers.}
\label{fig:toosmall}
\end{figure}
For more details we refer to \cite{WE96a} .

In Fig. 2 we compare the cost in molecular dynamics time for the Polyakov loop
operator as a function of stepsize for both algorithms. Data with $k=4$ and
$k=8$ are included and compared with HMC using a trajectory length $\tau_0=1$. 
We consider the pattern of this plot similar to the findings of \cite{HMC2}
in their figure 2.

\subsection{HMC on the CRAY T3D}

The computational demands of our study
let us move to
the massively parallel processing CRAY T3D
with a peak rate of  150 Mflops in 64 bit arithmetic per node.
We decided to implement the shared memory routines because they
show less overhead than Parallel Virtuell Machine or Message 
Passing Interface. We found the expected
speedup within a few percents for 4 PEs up to 128 PEs on a $16^4$ lattice.
For our present implementation of the HMC algorithm 5 PEs on a T3D are
equivalent to one YMP processor indicating the improvment of
the performance to cost ratio \cite{WE96b}.

\section{CONCLUSIONS}
We find that reasonable high statistics are absolutely necessary to obtain 
reliable error bars for the integrated autocorrelation times.
We consider the different decay behaviour of the integrated autocorrelation
functions for the observables studied so far as an indication for
different efficiency.
Up to now we did not find a competitive set of parameters for the
Kramers equation algorithm.

\section{ACKNOWLEDGEMENT}
One of us (S.M) enjoyed the kind hospitality of SCRI, Florida State
University and of IWR , University of Heidelberg while the work
described has been performed. It is a great pleasure to thank
Kahil Bitar, Urs Heller, Karl Jansen and in particular Tony
Kennedy for discussions and comments. We thank CRAY Research and
in particular Dr.R.~Fischer for support .

\end{document}